\documentclass[aps,prl,twocolumn,groupedaddress,showpacs]{revtex4}
\usepackage{graphicx}
\usepackage{dcolumn}
\usepackage{bm}

\def\ii{{\rm i}} \def\ee{{\rm e}}  \def\jb{{\bf j}}
\def\Eb{{\bf E}} \def\Rb{{\bf R}} \def\rb{{\bf r}}  
\def\zt{{\hat{\bf z}}}  
   \def\Hb{{\bf H}}
 \def\Ab{{\bf A}}

\def\kb{{\bf k}}

\begin{document}
\title{Diacritical study of light, electrons, and sound scattering by particles and holes}
\author{F.~Javier~Garc\'{i}a~de~Abajo,$^{1,3}$ H\'ector~Estrada,$^{2}$ and Francisco~Meseguer$^{2}$}

\affiliation{$^1$Instituto de \'Optica - CSIC, Serrano 121, 28006 Madrid, Spain \\
$^2$Unidad Asociada ICMM - CSIC/Universidad Polit\'ecnica de Valencia, Av. de los Naranjos s/n, 46022 Valencia, Spain \\
$^3$Author to whom any correspondence should be addressed: jga@cfmac.csic.es}

\begin{abstract}
We discuss the differences and similarities in the interaction of scalar and vector wave-fields with particles and holes. Analytical results are provided for the transmission of isolated and arrayed small holes as well as surface modes in hole arrays for light, electrons, and sound. In contrast to the optical case, small-hole arrays in perforated perfect screens cannot produce acoustic or electronic surface-bound states. However, unlike electrons and light, sound is transmitted through individual holes approximately in proportion to their area, regardless their size. We discuss these issues with a systematic analysis that allows exploring both common properties and unique behavior in wave phenomena for different material realizations.
\end{abstract}

  \maketitle

\section{Introduction}

Scientists have been fascinated by wave phenomena and the many common aspects shared by different physical realizations of propagating oscillatory motion, although unique behavior is also accompanying each of them \cite{B1953}. Light, electrons, and sound are prototypical examples of waves that reveal profound differences, even at the level of a single-particle, classical description. This discussion is preceded by a rich literature dating back to the nineteenth century \cite{R1897,R1897_2,B1941,B1944,LS1948,S1948,B1953,B1954,WS1965}. Atom and electron waves are at the heart of a long series of studies in thermal-atom scattering \cite{FR98} and electron diffraction by atomic lattices \cite{paper060}, whereas the discovery of extraordinary light transmission through hole arrays \cite{ELG98} has also prompted new investigations of sound \cite{HMK07,LLF07,paper154,CMG08,paper169} and reactivated the field.



The similarities between acoustic and optical transmission have been emphasized by several groups \cite{HMK07,LLF07,CMG08}, but intrinsic differences separate the two kinds of waves, namely: the absence of a cutoff wavelength for the existence of acoustic guided modes, in contrast to the optical case; the extraordinary shielding of sound near the onset of diffraction \cite{paper154}; and the exotic behavior conveyed by intrinsic elastic modes, with no parallel in optics \cite{paper169}. Likewise, matter waves display unique characteristics, as we discuss below. However, despite the intensity of the debate, no concluding study has been made on the similarities and differences in the scattering of scalar and vector waves by particles and holes, although the accepted emphasis on similarities is a common source of misconceptions.

Here, we confront the scattering properties of particles and holes for different kinds of waves (light, electrons, and sound). First, we recall that small individual holes drilled in perfect screens are very permeable to sound, but nearly impenetrable for electrons and light.  We then revisit the Babinet principle for all these waves, which directly relates the properties of perforated screens to those of particles arrays. From here, we demonstrate that hole arrays in thin screens cannot sustain acoustic or electronic surface-trapped modes, in contrast to the optical case. These results provide essential tools for exploring other common properties and unique behavior in wave phenomena, some of which are outlined in the concluding paragraph.

The electron wave function, the sound pressure in a fluid, and each of the components of the electromagnetic field satisfy the scalar wave equation
\begin{equation}
(\nabla^2+k^2)\psi=0,
\label{waveeq}
\end{equation}
where $k$ is the wavevector. Fluid dynamics, Schr\"oginger's equation, and Maxwell's equations reduce to Eq.\ (\ref{waveeq}) when describing sound, electrons, and light in homogeneous media. The difference in the behavior of these types of waves lies in the boundary conditions. For simplicity, we assume infinite potential boundaries for electrons, hard solids for sound, and perfect conductors for light. Then, the fields at the boundaries satisfy $\psi=0$ for electrons, the vanishing of the normal derivative for sound, and the vanishing of the parallel electric field and the perpendicular magnetic field for light.

\begin{figure*}
\center \includegraphics[width=160mm]{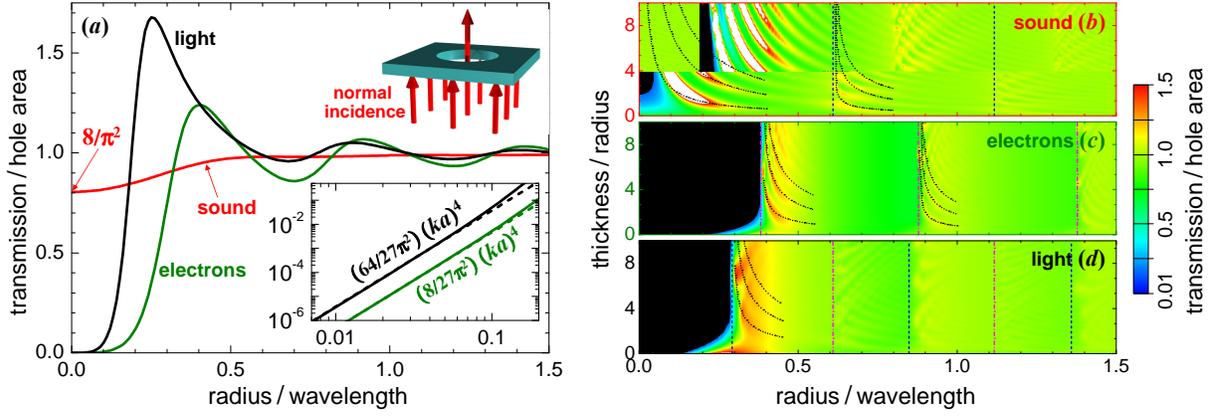}
\caption{\label{Fig1} Transmission of light, electrons, and sound through an individual circular hole, drilled in perfect-conductor, infinite-potential, and hard-solid films, respectively. {\bf (a)} Transmission cross section of a zero-thickness screen as a function of aperture radius. The transmission is normalized to the hole area. The lower inset shows the small-hole behavior (dashed curves and analytical expressions) compared to full calculations (solid curves). The small-hole analytical limits are taken from Refs. \cite{B1941}, \cite{B1944}, and \cite{LS1948} for sound, light, and electrons, respectively. {\bf (b-d)} Full thickness-radius dependence of the normalized transmission. The onsets of new guided modes are indicated by vertical dashed lines. The analytical estimates for Fabry-Perot resonances are shown as dotted curves.}
\end{figure*}

\begin{table*}
\begin{center}
\begin{tabular}{c|c|cl} \hline type of wave & scattered field & scattering coefficients & \\ \hline
&&&\\
sound & $\psi^{\rm scat}(\rb)=\alpha_s\psi_0\frac{\ee^{\ii kr}}{r}$ & ${\rm Re}\{\alpha_s\}=-a/\pi$ & ${\rm Im}\{\alpha_s^{-1}\}=-2k$
\\&&&\\
electrons & $\psi^{\rm scat}(\rb)=\alpha_e(\ii\frac{\partial\psi_0}{\partial z})\frac{\partial}{\partial z}(\frac{\ee^{\ii kr}}{r})$ & ${\rm Re}\{\alpha_e\}=-a^3/3\pi$ & ${\rm Im}\{\alpha_e^{-1}\}=-2k^3/3$
\\&&&\\
light & $\Hb^{\rm scat}(\rb)=\big[\alpha_M\left(k^2\Hb_0+\Hb_0\cdot\nabla\frac{\partial}{\partial z}\right)$ & ${\rm Re}\{\alpha_M\}=$
& ${\rm Im}\{\alpha_M^{-1}\}=$
\\& $+\ii k\alpha_E\Eb_0\times\nabla\big]\frac{\ee^{\ii kr}}{r}$ & $-2{\rm Re}\{\alpha_E\}=2a^3/3\pi$ & ${\rm Im}\{\alpha_E^{-1}\}=-2k^3/3$
\\&&&\\
\hline
\end{tabular}
\end{center}
\caption{Near-side scattered field produced by an individual aperture of radius $a$ drilled in a thin perfect screen in response to a scalar field $\psi_0$ (for sound and electrons) or perpendicular-electric and parallel-magnetic fields $\Eb_0$ and $\Hb_0$ (for light) acting at the position of the hole. The transmitted field on the far side has the same form, but with opposite values of the scattering coefficients $\alpha_s$, $\alpha_e$, $\alpha_E$, and $\alpha_M$. Notice that the scattered field is constructed from the solution of Eq.\ (\ref{waveeq}) for $r\neq 0$, $\exp(\ii kr)/r$, via these scattering coefficients. In particular, $\alpha_E$ and $\alpha_M$ are the electric and magnetic polarizabilities \cite{J99}, respectively.}
\label{table1}
\end{table*}

\section{Individual holes}

The response of an individual hole drilled in a thin perfect screen reveals distinct characteristics for different types of waves, as illustrated in Fig.\ \ref{Fig1} and Table\ \ref{table1}. The transmission in Fig.\ \ref{Fig1} is calculated by matching continuous-cylindrical-wave expansions of the fields outside the hole to a sum of guided modes inside the cylindrical cavity \cite{R1987,JE95}. The latter reduce to $J_m(QR)\exp(\pm\ii qz+\ii m\varphi)$ in the case of scalar waves (expressed in cylindrical coordinates $(R,\varphi,z)$), where $m$ is the azimuthal number, $Q$ runs over the zeros of $J_m(Qa)$ and $J'_m(Qa)$ for electrons and sound, respectively, $a$ is the hole radius, and $q=\sqrt{k^2-Q^2}$ is the wavevector along the hole-axis direction, $z$. In particular, only $m=0$ contributes for sound ($Qa=0,\,3.83,\,7.02,\dots$) and electrons ($Qa=2.40,\,5.52,\cdots$) under normal incidence. The hole-cavity optical modes possess a polarization degree of freedom \cite{R1987}, so that doubly-degenerate $m=\pm 1$ waves participate in normal-incidence scattering ($Qa=1.84,\,3.83,\,5.33,\cdots$).

The transmission exhibits the following interesting features: (1) Holes in a zero-thickness screen [Fig.\ \ref{Fig1}(a)] transmit sound approximately in proportion to their area \cite{B1941}, in contrast to the severe $\sim(a/\lambda)^4$ cutoff for light \cite{B1944} and electrons \cite{LS1948}. (2) The transmission is enhanced and displays a colorful structure near the onset of new guided modes (i.e., near $k=Q$), as indicated in Fig.\ \ref{Fig1}(b) by vertical dashed lines. (3) Fabry-Perot resonances are established upon successive reflections of guided modes at both hole ends, signalled by the in-phase scattering condition $qt=n\pi+\varphi_0$, where $t$ is the thickness, $n$ runs over positive integers, $q$ depends on $k$ as noted above, and $\varphi_0$ is a phase shift produced by the noted reflections (we fit $\varphi_0=0,\,-\pi/4$, and $-\pi/2$ for light, electrons, and sound, respectively). This condition leads to the dotted curves in Fig.\ \ref{Fig1}(b) (i.e., a relation between $t$ and $k$), in reasonable agreement with the calculated transmission maxima, although they do not exactly map the complexity of modal structures arising from mode interference. Also, good correspondence between measurements and simulations was reported for these acoustic resonances by Wilson and Soroka as early as 1965 \cite{WS1965}. (4) The acoustic scattered pressure is a monopole [$\exp(\ii kr)/r$], in contrast to the dipolar response to electrons and light, as inferred from Table\ \ref{table1} (second column), which shows small-aperture close-form expressions derived from an analysis similar to that of Ref.\ \cite{J99} for light. This analysis yields an overall multiplicative scattering coefficient, $\alpha_\nu$ (Table\ \ref{table1}, third column), the imaginary part of which is found by imposing the condition that the scattered and reflected waves carry the same flux as an incident plane wave, because there is no absorption in a perfect screen. Then, a non-vanishing ${\rm Im}\{\alpha_\nu\}$ reflects the finite coupling of the hole to propagating waves. Incidentally, the small-hole transmission cross sections given in the inset of Fig.\ \ref{Fig1}(a) follow from angular integration of the far-field intensity \cite{B1941,B1944,LS1948}.

\begin{table*}
\begin{center}
\begin{tabular}{c|c|c|c} \hline type of wave & lattice sum & density of states & transmission coefficient \\ \hline
&&&\\
sound & $G_s=\sum_{\Rb\neq 0}\ee^{-\ii\kb_\parallel\cdot\Rb}\frac{\ee^{\ii kr}}{r}$ & ${\rm Im}\{G_s\}=\frac{2\pi L}{k_z}-k$
& $t_s=[L+\ii(Ak_z/2\pi){\rm Re}\{\alpha_s^{-1}-G_s\}]^{-1}$
\\&&&\\
electrons & $G_e=\sum_{\Rb\neq 0}\ee^{-\ii\kb_\parallel\cdot\Rb}\frac{\partial^2}{\partial z^2}\frac{\ee^{\ii kr}}{r}$ & ${\rm Im}\{G_e\}=\frac{-2\pi k_zL}{A}+\frac{k^3}{3}$
& $t_e=-[L-\ii(A/2\pi k_z){\rm Re}\{\alpha_e^{-1}-G_e\}]^{-1}$
\\&&&\\
light & $G_l=\sum_{\Rb\neq 0}\ee^{-\ii\kb_\parallel\cdot\Rb}[k_z^2-(\hat{\kb}_\parallel\cdot\nabla)^2]\frac{\ee^{\ii kr}}{r}$ & ${\rm Im}\{G_l\}=\frac{2\pi k_zL}{A}-\frac{2k^3}{3}$
& $t_{\rm TE}=[L+\ii(Ak_z/2\pi){\rm Re}\{\alpha_M^{-1}-G_l\}]^{-1}$
\\&&&\\
\hline
\end{tabular}
\end{center}
\caption{Lattice sums and transmission coefficients for periodic small-hole arrays. The parallel wavevector of the incident plane wave is $\kb_\parallel$, the normal wavevector is $k_z=\sqrt{k^2-k^2_\parallel}$, and the sums run over lattice sites $\Rb$. $A$ is the area of the unit cell. The value of $L$ is 1 in region I of Fig.\ \ref{Fig2}(a) (i.e., above $k=k_\parallel$ and below the diffraction threshold) and 0 in region II (i.e., below both $k=k_\parallel$ and the diffraction threshold).}
\label{table2}
\end{table*}

\begin{figure}
\center \includegraphics[width=80mm]{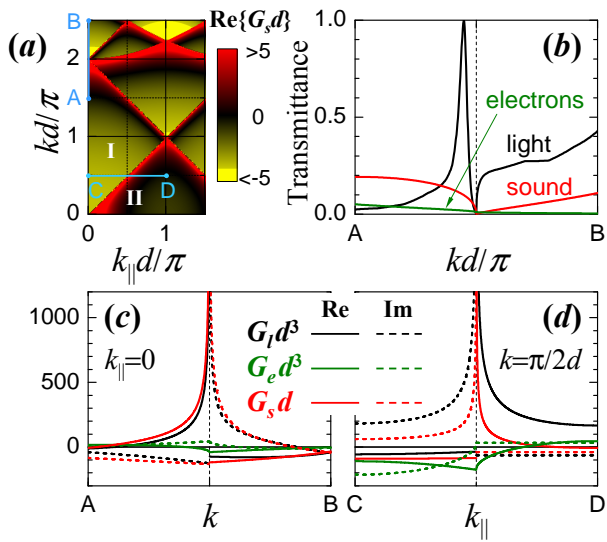}
\caption{\label{Fig2} Interaction between holes in a zero-thickness screen. We assume a square lattice of period $d$. {\bf (a)} Real part of the lattice sum $G_s$ for interaction of sound scattered by holes in a screen. The parallel wavevector $\kb_\parallel$ is taken along a nearest-neighbors direction. {\bf (b)} Normal incidence transmission of light, electrons, and sound through holes of radius $d/5$ for an excursion of the wavevector $k$ along the AB segment in the diagram of (a). {\bf (c)} Lattice sums for all three kinds of waves (light $G_l$, electrons $G_e$, and sound $G_s$) along the AB segment of (a). Real and imaginary parts are represented as solid and broken curves, respectively. {\bf (d)} Same as (c), but along the CD segment of (a). The $G_s$ curve in (c) and all curves in (d) are multiplied by a factor of 25.}
\end{figure}

\section{Transmission through hole arrays}

The response of a periodic hole array to an incident plane wave can be readily expressed in terms of the response of each individual element (see scattering coefficients in Table\ \ref{table1}) both to the external perturbation and to the field produced by the rest of the holes \cite{paper130}. The latter reduces to a sum of the inter-hole interaction over the lattice, $G_\nu(\kb_\parallel)$, which must include a phase factor $\exp(-\ii\kb_\parallel\cdot\Rb)$ depending on the lattice site $\Rb$ and driven by the incident-plane-wave parallel-wavevector $\kb_\parallel$, as shown in the explicit expressions given in Table\ \ref{table2} (second column). Interestingly, ${\rm Im}\{G_\nu\}$ admits close-form expressions (Table\ \ref{table2}, third column) because only a finite number of diffracted beams contribute when re-expressing $G_\nu$ as a sum over reciprocal lattice vectors. In particular, the first term in ${\rm Im}\{G_\nu\}$ vanishes ($L=0$) in region II of Fig. \ref{Fig2}(a) (see Table\ \ref{table2}).

Following a procedure detailed elsewhere for the case of light \cite{paper130}, we find the transmission coefficients $t_\nu$ listed in Table\ \ref{table2} (last column). In the optical case, we just give the transmission for TE polarization in the hole array, but a similar expression is obtained for TM transmission. The $t_\nu$ coefficients have been simplified by using the analytical expressions for ${\rm Im}\{\alpha_\nu^{-1}\}$ and ${\rm Im}\{G_\nu\}$ given in Tables\ \ref{table1} and \ref{table2}. The remaining real part ${\rm Re}\{\alpha_\nu^{-1}-G_\nu\}$ appears in the denominator of $t_\nu$ and full transmission is predicted when it vanishes within region I of Fig.\ \ref{Fig2}(a), where $L=1$, so that $|t_\nu|=1$ under that condition.

In particular, small holes produce large values of $\alpha_\nu^{-1}$ that can be only matched by divergences in $G_\nu$ such as those shown in Fig.\ \ref{Fig2}(a) for $G_s$. For example, ${\rm Re}\{G_s\},\,{\rm Re}\{G_l\}\rightarrow\infty$ along the $k=k_\parallel$ and $k=2\pi/d-k_\parallel$ lines. This is illustrated in Fig.\ \ref{Fig2}(c) for $k_\parallel=0$. However, ${\rm Re}\{\alpha_s^{-1}\}<0$, so that sound cannot be fully transmitted through small holes (i.e., ${\rm Re}\{G_s\}>0$ and ${\rm Re}\{\alpha_s\}<0$ have opposite signs in the region where ${\rm Re}\{G_s\}$ diverges, so that ${\rm Re}\{\alpha_\nu^{-1}-G_\nu\}$ cannot vanish). In contrast, 100\% transmission of light is possible for arbitrarily-small holes because ${\rm Re}\{\alpha_M\}>0$, and ${\rm Re}\{G_l\}$ can always be matched by ${\rm Re}\{\alpha_M\}{\mathop{\longrightarrow}\limits_{a\to 0}}\infty$ \cite{paper130}. Regarding electrons, $G_e$ does not even diverge, thus averting the possibility of full matter-wave transmission through small holes.

Additionally, a divergence in $G_\nu$ results in vanishing transmission, and this is the case of light and sound for $k=2\pi/d-k_\parallel$, in contrast to electrons, which have finite transmission under that condition [see, for example, the midpoint between A and B in Fig.\ \ref{Fig2}(b)].

Although these conclusions are predicted by the small-aperture limits discussed in Table\ \ref{table2}, they are in full agreement with the rigorous calculations shown in Fig.\ \ref{Fig2}(b), which are obtained by matching expansions of the field similar to those employed for individual holes (in particular, Rayleigh expansions outside the film, which reflect the periodicity of the array).

\section{Bound states in hole arrays}

Trapped modes are signalled by divergences in the transmission coefficients outside the diffraction region (i.e., when there is scattered evanescent field without incident waves). This is possible in region II of Fig. \ref{Fig2}(a), where $L=0$, so that $t_\nu$ diverges if ${\rm Re}\{\alpha_\nu^{-1}-G_\nu\}=0$ (see Table\ \ref{table2}, last column). This possibility is explored in Fig.\ \ref{Fig2}(d), which shows the lattice sums $G_\nu$ along the CD segment of Fig.\ \ref{Fig2}(a). Similar to the transmission discussed above, small holes can meet this condition provided ${\rm Re}\{G_\nu\}$ diverges and ${\rm Re}\{\alpha_\nu\}$ is positive. This can be only true for light, and therefore, we predict optical bound states in arrays of holes, however small. In contrast, no acoustical or electronic modes are possible in such arrays made of regular, small holes. These conclusions are summarized in Table\ \ref{table3}.

\section{Particle arrays}

Invoking Babinet's principle, the transmission of sound, electrons, and light through a hole array drilled in a thin screen must coincide with the reflection of electrons, sound, and light (in that order), on the complementary array of disks (see Appendix\ A). Therefore, an array of small disks can produce full reflection, but this is not the case for sound and electrons. In contrast to previous studies \cite{HMK07}, the same argument leads to the conclusion that no acoustic or electronic modes are possible in arrays of small holes or disks for thin specimens, although TE (TM) optical modes exist in disk (hole) arrays \cite{paper130}. Further inspection of the scattering coefficients for small spherical scatterers made of infinite-potential and hard-solid materials reveals that planar periodic arrays composed of such objects cannot trap electrons or sound, respectively, although finite potentials and softer materials can lead to both full reflection and planar binding via localized particle modes.

\begin{table}
\begin{center}
\begin{tabular}{c|c|c|c} \hline type of wave & \,\,\,one-hole\,\,\, & \,\,\,hole-array\,\,\, & hole-array
\\ & ET & ET & \,\,\,bound states\,\,\,
\\ \hline&&&\\
sound & Yes & No & No
\\&&&\\
electrons & No & No & No
\\&&&\\
light & No & Yes & Yes
\\&&&\\
\hline
\end{tabular}
\end{center}
\caption{Extraordinary transmission (ET) of individual and arrayed small holes, and surface-bound modes in arrays. The acoustic transmission of individual holes is proportional to their area, whereas arbitrarily-small-hole arrays can trap light and produce full optical transmission.}
\label{table3}
\end{table}

\section{Summary}

The scattering of sound, electrons, and light by arrays of small particles and holes exhibit dramatic differences that are clearly illustrated by Fig.\ \ref{Fig1}(a) and by the existence of surface modes exclusively for light (see a summary in Table\ \ref{table3}). Beyond the scope of our work, we expect the effect of non-perfect screens and multipolar scattered terms to also produce exotic behavior, in particular when non-absorbing scatterers are considered (for example, real potentials and dielectric materials for electrons and light). Acoustic and electron scatterers can be made to resonate through their monopolar or dipolar response, thus leading to new classes of surface modes in arrays. We hope that our study stimulates further theoretical and experimental investigation in these directions.

\section*{Acknowledgement}

This work has been supported by the Spanish MICINN (MAT2006-03097, MAT2007-66050, and Consolider NanoLight.es) and the EU (NMP4-2006-016881-SPANS and NMP4-SL-2008-213669-ENSEMBLE). H.E. acknowledges a CSIC-JAE predoc scholarship.


\appendix

\section*{Appendix A. Babinet's principle for light, electrons, and sound}

In this appendix, we present a simple, self-contained derivation of the Babinet principle for scalar and vector waves, which directly relates the properties of any perforated thin planar screen to those of its complementary screen, thus allowing us to extract conclusions on the relation between particle and hole arrays regarding their transmission, reflection, and wave binding performance. The sound pressure reflected by the former screen coincides with the forward scattering wave function for electrons incident on the complementary screen. For light, the field reflected by the former screen for a given polarization of the incident wave coincides with the field transmitted by the latter screen for the orthogonal polarization.

The waves scattered by a thin perfect screen and by its complementary screen are related via Babinet's principle. By perfect screen we understand an infinite-potential wall for electrons, a hard-solid plate for sound, or a perfect-conductor film for light.
The evolution of these waves in a homogeneous medium is ruled by the same scalar wave equation,
\begin{equation}
(\nabla^2+k^2)\psi=0,
\label{waveeq}
\end{equation}
where $k$ is the wavevector, and $\psi$ is the wave function for electrons, the pressure for sound, and the components of the electromagnetic field for light. The fields at the boundaries satisfy $\psi=0$ for electrons, the vanishing of the normal derivative for sound, and the cancelation of the parallel electric field and the perpendicular magnetic field for light.

\begin{figure*}
\center \includegraphics[width=100mm]{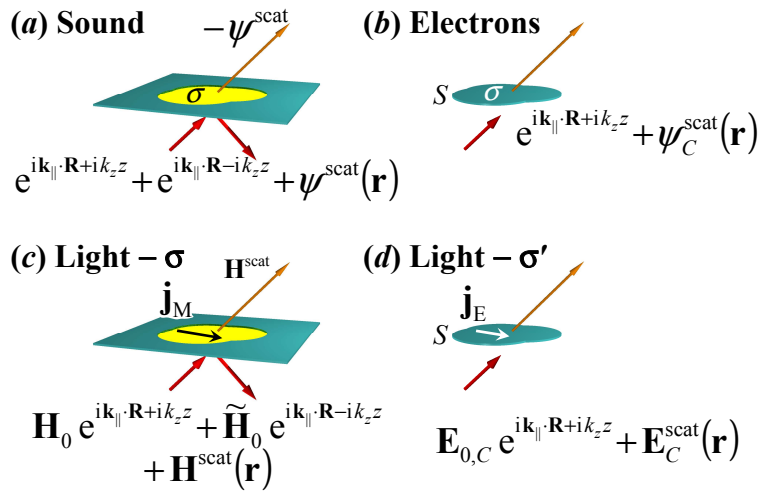}
\caption{\label{FigAM} Elements used in the discussion of Babinet's principle, which connects the scattering of sound by a screen containing an aperture region $S$ (a) and the scattering of electrons by the complementary screen (b), so that $\psi^{\rm scat}=\psi_C^{\rm scat}$ is shared in both cases. It also connects the scattering of light with a given polarization $\sigma$ (e.g., $\sigma=$TE or TM) (c) and the scattering of light with orthogonal polarization $\sigma'$ (i.e., $\sigma'=$TM or TE, respectively, and also $\Eb_{0,C}=\Hb_0$) by the complementary screen (d), so that the scattered fields in (c) and (d) satisfy $\Eb_C^{\rm scat}=\Hb^{\rm scat}$. The screens are contained in the $z=0$ plane.}
\end{figure*}

\subsection*{Babinet's principle for scalar waves}

The properties of a patterned thin screen (zero thickness) and those of the complementary screen are related via Babinet's principle \cite{R1897,R1897_2}. For an external sound plane-wave, $\exp(\ii\kb_\parallel\cdot\Rb+\ii k_zz)$, incident on the first screen [Fig.\ \ref{FigAM}(a)], we can fulfill the condition of vanishing normal derivative by just adding a specularly reflected wave, $\exp(\ii\kb_\parallel\cdot\Rb-\ii k_zz)$, where $\Rb$ is the parallel component of the coordinate vector $\rb=(\Rb,z)$, and the direction $z$ is normal to the screen. We still need to impose the continuity of both the wave amplitude and its normal derivative at the screen opening $S$, which can be done by simply adding the following scattered wave:
\begin{equation}
\psi^{\rm scat}(\rb)={\rm sign}(z)\int_Sd\Rb'\;\mathcal{G}^0(\rb-\Rb')\;\sigma(\Rb'),
\label{psiscat}
\end{equation}
where
\begin{equation}
\mathcal{G}^0(r)=\exp(\ii kr)/r
\nonumber
\end{equation}
is the Green function of the wave equation [Eq.\ (\ref{waveeq})] and $\sigma$ is a {\it surface charge} defined on $S$ and determined by the condition
\begin{equation}
\psi^{\rm scat}(\Rb,z=0)=-1,\;\;\;\;\Rb\in S.
\nonumber
\end{equation}
Noticing that
\begin{equation}
\partial\mathcal{G}^0(\rb)/\partial z{\mathop{\longrightarrow}\limits_{z\to\pm 0}}\mp\delta(\Rb),
\label{eqder}
\end{equation}
we find that (i) the expression (\ref{psiscat}) has vanishing normal derivative in the screen outside $S$, thus satisfying the boundary conditions for Fig.\ \ref{FigAM}(a), and (ii) $\psi^{\rm scat}(\rb)+\exp(\ii\kb_\parallel\cdot\Rb+\ii k_zz)$ vanishes at $S$, so that it represents the solution for an electron wave function incident on the complementary screen [Fig.\ \ref{FigAM}(b)]. This means that the transmittance of sound (electrons) through a drilled screen coincides with the reflectance of electrons (sound) on the complementary disk(s).

\subsection*{Babinet's principle for light}

Similarly, the transmittance of light with a certain polarization through a perforated screen coincides with the reflectance of light with orthogonal polarization on the complementary screen \cite{paper130}. This statement of Babinet's principle is a rigorous consequence of vector diffraction theory \cite{J99}, but it admits a simpler derivation based upon the symmetry of Maxwell's equations \cite{HKL1955}, the solution of which reads, for external electric and magnetic currents $\jb_E$ and $\jb_H$,
\begin{equation}
\Eb^{\rm scat}=\frac{\ii}{k}(k^2+\nabla\nabla\cdot)\Ab_E-\nabla\times\Ab_H
\label{Escat}
\end{equation}
and
\begin{equation}
\Hb^{\rm scat}=\frac{\ii}{k}(k^2+\nabla\nabla\cdot)\Ab_H+\nabla\times\Ab_E,
\label{Hscat}
\end{equation}
where
\begin{equation}
\Ab_\nu=\frac{1}{c}\int_S d\Rb'\;\mathcal{G}^0(\rb-\Rb')\;\jb_\nu(\Rb').
\nonumber
\end{equation}
In our case, the currents are considered to be in the plane of the screen, within a region $S$ defining an opening, or a disk in the complementary geometry. In particular, the disk of Fig.\ \ref{FigAM}(d) displays an electric current $\jb_E$ (but no magnetic current) in response to an external plane wave of polarization $\sigma'$. Using the limit (\ref{eqder}) for $\partial\mathcal{G}^0/\partial z$, we find that the rotational terms in Eqs.\ (\ref{Escat}) and (\ref{Hscat}) are perpendicular to the screen outside $S$, and the remaining terms are parallel to it. This, together with the invariance of the equations under the transformation
\begin{eqnarray}
\jb_E&\rightarrow&\jb_H, \nonumber \\
\jb_H&\rightarrow&-\jb_E, \nonumber \\
\Eb&\rightarrow&\Hb, \nonumber \\
\Hb&\rightarrow&-\Eb, \nonumber
\end{eqnarray}
permits writing the solution for the complementary drilled screen of Fig.\ \ref{FigAM}(c), illuminated by a plane wave of orthogonal polarization $\sigma$ (for example, $\sigma=$TE and $\sigma'=$TM), using the scattered electric field of Fig.\ \ref{FigAM}(d) ($\Eb_C^{\rm scat}$), now taken to be the scattered magnetic field ($\Hb^{\rm scat}=\Eb_C^{\rm scat}$). Notice that the plane wave reflected by an unperforated film (i.e., $\tilde{\Hb}_0$ with $\tilde{\Hb}_{0\parallel}=\Hb_{0\parallel}$ and $\tilde{\Hb}_{0\perp}=-\Hb_{0\perp}$) has been added to Fig.\ \ref{FigAM}(c) in order to match the boundary conditions in the film. Direct inspection reveals that the condition
\begin{equation}
\left[\Eb_0\ee^{\ii\kb_\parallel\cdot\Rb}+\Eb_C^{\rm scat}(\Rb,z=0)\right]\times\zt=0,\;\;\;\;\Rb\in S.
\nonumber
\end{equation}
guarantees both the vanishing of $\Eb_\parallel$ and $\Hb_\perp$ in the disk of Fig.\ \ref{FigAM}(d) and the continuity of the fields in the hole of Fig.\ \ref{FigAM}(c).

\section*{References}


\end{document}